\documentclass[sigconf]{acmart}
\AtBeginDocument{%
 }

\usepackage{multirow}
\usepackage{siunitx}
\newcommand{\commentout}[1]{}

\usepackage{enumitem}
\setlist[description]{leftmargin=\parindent,labelindent=0pt}
\setlist[itemize]{leftmargin=\parindent,labelindent=0pt}
\usepackage{ulem} 

\setcopyright{acmlicensed}

\copyrightyear{2026}
\acmYear{2026}
\setcopyright{cc}
\setcctype{by}
\acmConference[CHI '26]{Proceedings of the 2026 CHI Conference on Human Factors in Computing Systems}{April 13--17, 2026}{Barcelona, Spain}
\acmBooktitle{Proceedings of the 2026 CHI Conference on Human Factors in Computing Systems (CHI '26), April 13--17, 2026, Barcelona, Spain}
\acmDOI{10.1145/3772318.3791397}
\acmISBN{979-8-4007-2278-3/2026/04}

\begin{document}

\title{NasoVoce: A Nose-Mounted Low-Audibility Speech Interface for Always-Available Speech Interaction}

\author{Jun Rekimoto}
\authornote{Corresponding author.}
\orcid{0000-0002-3629-2514}
\affiliation{%
\institution{Sony CSL - Kyoto}
\city{Kyoto-shi}
\state{Kyoto}
\country{Japan}
}
\affiliation{%
 \institution{The University of Tokyo}
\city{Bunkyo-ku}
\state{Tokyo}
\country{Japan}
\postcode{113-0033}
}
\email{rekimoto@acm.org}

\author{Yu Nishimura}
\orcid{0009-0004-3978-6917}
\affiliation{%
 \institution{Sony CSL}
\city{Shinagawa-ku}
\state{Tokyo}
\country{Japan}
\postcode{141-0022}
}
\email{nishimura@csl.sony.co.jp}

\author{Bojian Yang}
\orcid{0000-0001-9378-2555}
\affiliation{%
 \institution{Sony CSL}
\city{Shinagawa-ku}
\state{Tokyo}
\country{Japan}
\postcode{141-0022}
}
\email{bojian.yang@sony.com}

\renewcommand{\shortauthors}{Rekimoto et al.}

\begin{abstract}
Silent and whispered speech offer promise for always-available voice interaction with AI, yet existing methods struggle to balance vocabulary size, wearability, silence, and noise robustness. We present NasoVoce, a nose-bridge–mounted interface that integrates a microphone and a vibration sensor. Positioned at the nasal pads of smart glasses, it unobtrusively captures both acoustic and vibration signals. The nasal bridge, close to the mouth, allows access to bone- and skin-conducted speech and enables reliable capture of low-volume utterances such as whispered speech. While the microphone captures high-quality audio, it is highly sensitive to environmental noise. Conversely, the vibration sensor is robust to noise but yields lower signal quality. By fusing these complementary inputs, NasoVoce generates high-quality speech robust against interference. Evaluation with Whisper Large-v2, PESQ, STOI, and MUSHRA ratings confirms improved recognition and quality. NasoVoce demonstrates the feasibility of a practical interface for always-available, continuous, and discreet AI voice conversations.
\end{abstract}




\begin{teaserfigure}
\centering
 \includegraphics[width=0.88\textwidth]{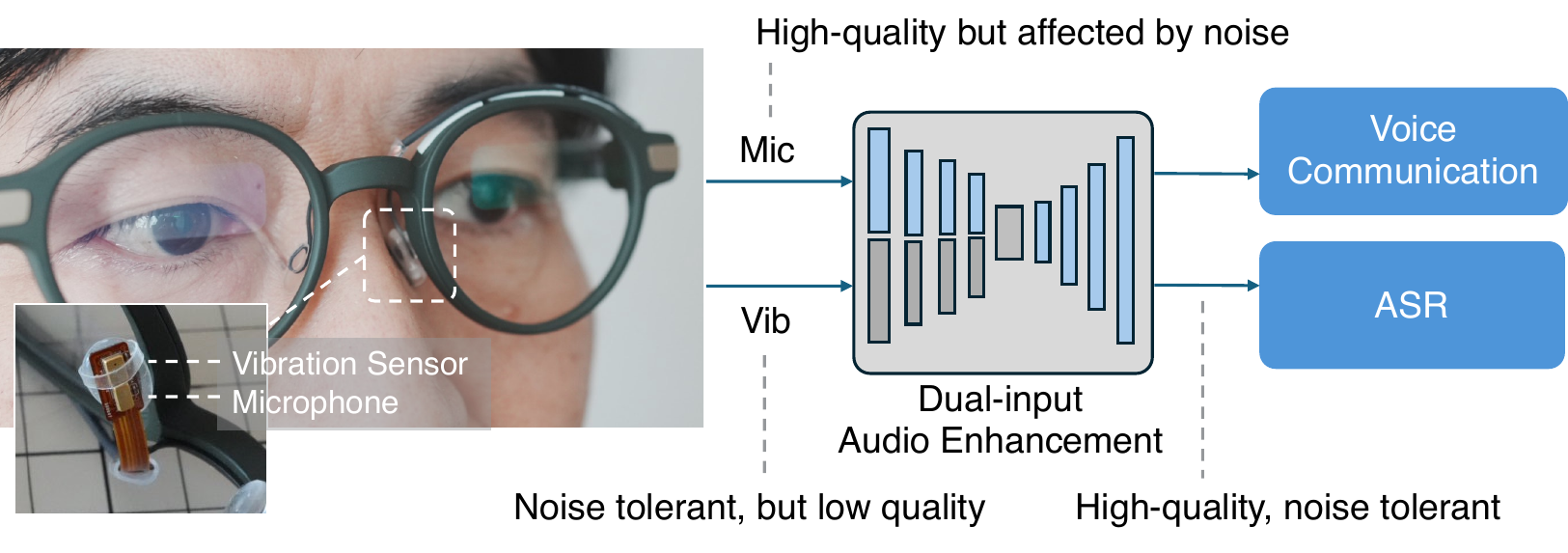}
 \caption{NasoVoce integrates a microphone and a vibration sensor at the nasal bridge for capturing normal and whispered speech. This location enables effective capture of both air-conducted and bone/skin-conducted signals even in very noisy environment, while remaining unobtrusive for daily wear.}
 \label{fig:teaser}
\end{teaserfigure}

\maketitle

\section{Introduction}

In daily life, the usage of generative AI is increasing. Beyond supporting document writing and programming, generative AI is also effective as a tool for assisting thought processes, such as idea exploration and information retrieval. This function is particularly valuable in voice mode. For instance, users can record spontaneous ideas verbally, or query the AI for opinions and related information, enabling its use as an always-available voice agent. 

Recent advances in glasses-type devices, which integrate microphones, speakers, cameras, and in-lens displays, provide an ideal platform for such continuous interaction. The combination of voice and image input allows users to ask questions about surrounding objects or scenery, or inquire about the operation of nearby equipment, enabling context-aware conversations with AI in real-world settings.

However, voice-based interfaces face several challenges: social acceptability (speaking aloud in public), privacy (difficulty vocalizing confidential information in public), and noise robustness (avoiding interference from environmental noise or nearby speech). Wearability is also a significant issue.

Previous research on ``silent speech'' interfaces, which attempt to recognize speech utterances using sensors, has primarily been limited to small command sets and remains insufficient for open-ended AI conversation. Lip-reading techniques have an extended vocabulary size, but camera-based systems covering the face impose high wearability costs, limiting their suitability for daily use.

As a method related to silent speech, ``whispered speech'' has also been proposed. Input using sufficiently soft whispers shares characteristics with silent speech in that it minimizes disturbance to the surrounding environment. Moreover, by employing speech recognition models optimized for whispered speech, the system can support an effectively unrestricted vocabulary, offering a major advantage~\cite{10.1145/3544548.3580706,10.1145/3706599.3721185,farhadipour2024leveragingselfsupervisedmodelsautomatic,10.1145/3652920.3652925,10888480}. However, because whispered speech is inherently low in volume, it is highly susceptible to interference from ambient noise.

We identify the following requirements for a practical always-available voice interface:

\begin{description}
\item[Wearability:] Devices must impose minimal burden during daily and continuous use. In-ear canal devices, such as earbuds, cause discomfort over long durations, and noise-canceling earphones may isolate users from environmental sounds, leading to safety concerns. Approaches requiring surface mounted sensors (e.g., intraoral or EEG electrodes) or a lip-reading camera in front of the user's face, are also unsuitable for continuous everyday use.
\item[Silence:] User speech should not be externally audible, avoiding both disturbance to others in public and leakage of private information.
\item[Noise Robustness:] Recognition must remain accurate under environmental noise and avoid misrecognition of nearby speech.
\item[Vocabulary:] The system should support a vocabulary comparable to standard speech recognition. Silent speech systems restricted to dozens of commands are inadequate for general AI conversation, and multilingual support is desirable.
\item[Speech Rate:] Input should be possible at normal conversational speed. Letter-by-letter spelling or telegraphic input cannot support natural dialogue.

\end{description}

To the best of our knowledge, no existing system simultaneously satisfies all of these requirements, as summarized in Table~\ref{tab:comparison}. In this work, we propose {\it NasoVoce}, a silent speech interface that integrates a microphone and a vibration sensor into the nasal bridge. This anatomical location, close to the mouth and nasal cavity, is well suited to capture air-conducted sounds of both normal and whispered speech via a microphone. 
Unlike throat microphones or bone-conduction sensors placed on the skull which primarily rely on vocal cord vibrations (voiced signals), the nasal bridge possesses a unique acoustic property; it effectively captures the aerodynamic turbulence and cavity resonance generated even during unvoiced whispering. 
Simultaneously, speech vibrations can be acquired using a vibration sensor, enabling robust speech recognition against environmental noise. 
This advantage allows NasoVoce to detect whispered speech signals that are physically undetectable by conventional throat or skull-contact sensors, establishing a distinct advantage for silent voice interaction.

This configuration is also particularly well-suited for integration into smart glasses, which are expected to become increasingly widespread. Embedding the system in the nasal component of smart glasses allows them to function as silent AI communication devices.

\begin{table*}
\begin{center}
\begin{tabular}{l|cccc}

\toprule
name & method & silence & noise tolerance & command/speech \\
\midrule
AlterEgo~\cite{10.1145/3172944.3172977} & neuromuscular & silent & good & command \\
SottoVoce~\cite{sottovoce} & ultrasound & silent & good & limited vocab. \\
SilentVoice~\cite{10.1145/3242587.3242603} & ingressive speech    & low & unknown & speech \\
WESPER~\cite{10.1145/3544548.3580706} & mic & normal/whisper & unknown & speech \\
LipLerner~\cite{10.1145/3544548.3581465} & lip-reading & silent & good & command \\
V-Speech~\cite{10.1145/3287058} & BC on nosepad & normal voice & good & speech \\
VibeVoice~\cite{He2023_MobiSys_VibVoice} & mic/IMU on HMD & normal voice & good & speech \\
Unvoiced~\cite{10.1145/3666025.3699374} & jawbone motion $+$ LLM & silent & good & phrase \\
AccCall~\cite{10.1145/3749463} & mic $+$ smartphone IMU & normal & good & speech \\
EchoSpeech~\cite{10.1145/3544548.3580801} & ultrasound on eyeglasses frame & normal & good & command \\
AirPods~\cite{Apple2025VoiceIsolation} & beamforming/signal processing & normal & good & speech \\
{\bf NasoVoce (ours)} & {\bf mic/vibration on nose-pad} & {\bf normal/whisper} & {\bf good} & {\bf speech} \\
\bottomrule
\end{tabular}
\end{center}
\caption{Comparison of silent and whispered speech / wearable speech technologies}
\label{tab:comparison}
\end{table*}

The contributions of this research are:
\begin{itemize}
\item A whisper input mechanism by integrating a microphone and vibration sensor into nose pads; This design reduces ambient noise for both normal and whispered speech while maintaining the smart glasses' appearance and wearability.
\item Develop a deep learning model, constructed a dataset, and verified the actual effectiveness of the proposed method.
\end{itemize}

Unlike prior work on bone conduction and microphone input, this study focuses on whispered speech. We emphasize this modality to enable AI conversation in any situation, similar to silent speech.

\section{Related Work}

This study proposes a system capable of capturing normal and whispered speech that can be integrated into the nose pads of eyewear. In this context, we review related work from four perspectives—silent speech, whispered speech recording, speech recording via bone and skin conduction, and the fusion processing of air- and bone-conducted speech—and compare them with our proposed approach.

\subsection*{Silent Speech} 

Silent Speech Interfaces (SSIs) enable communication when an acoustic signal is unavailable or undesirable by decoding articulatory, neuromuscular, or other biosignals instead of air-borne speech~\cite{decodingSSI2025}.
Various sensing families have been investigated including surface electromyography (sEMG)~\cite{10.1145/3172944.3172977}, articulatory imaging (ultrasound tongue imaging and lip video), contactless RF, and neural pathways—while outlining persistent challenges such as speaker/session variability, latency, and deployment ergonomics. 

sEMG (neuromuscular) systems, such as AlterEgo~\cite{10.1145/3172944.3172977}, capture articulator muscle activity from the face/neck to recognize or resynthesize speech. Classical systems established feasibility and highlighted mode mismatch between audible vs. silent articulation; recent deep models improve ``voicing'' silent EMG and sequence-to-sequence resynthesis, yet cross-session robustness remains a core problem. 

Ultrasound tongue imaging (UTI), often paired with lip video, supports articulatory to acoustic mapping and silent recognition. SottoVoce demonstrated the feasibility of connecting existing voice-aware devices~\cite{10.1145/3544548.3580706}. STN modules refine continuous vocoding, adaptation and multi-speaker recognition under silent vs. modal mismatches. 
EchoSpeech employs multiple pairs of ultrasonic transducers and receivers on a glasses-type device to estimate speech based on cheek skin deformations during silent articulation~\cite{10.1145/3544548.3580801}. While it shares our goal of being integratable into a glasses form factor, it is restricted to recognizing 31 speech commands and does not function as a general speech interface.

Unvoiced senses jaw motion during silent speech using an Inertial Measurement Unit (IMU)~\cite{10.1145/3666025.3699374}. This system relies on the jaw (a secondary articulator), which suffers from inherent physical ambiguity where multiple sounds map to the same movement (one-to-many mapping). Consequently, it has the limitation of being unable to physically distinguish words with similar jaw motion profiles without strictly relying on LLM-based contextual inference to resolve the ambiguity.

Deep end-to-end lip-reading (for example, LipNet~\cite{assael2016lipnetendtoendsentencelevellipreading} and Liplearner~\cite{10.1145/3544548.3581465}) decodes speech act from images of the lips or mouth.
Lip-reading by a depth image sensor uses depth information of the speech act~\cite{10.1145/3613904.3642092}. 
The main challenges of lip-reading are the need to position a camera where it can capture the lips and the high sensitivity of the system to ambient lighting, making mobile use difficult. At the same time, as lip-reading technology advances, a new concern may emerge: even when users input speech through other silent speech methods, the capture of lip images could undermine privacy and confidentiality.

\subsection*{Whispered Speech}

Although not completely silent, unvoiced whispering—produced without vocal cord vibration—possesses characteristics similar to silent speech, particularly in its minimal impact on the surrounding environment. Unlike silent speech approaches, whispered speech recognition enables performance comparable to general ASR with an unconstrained vocabulary. WESPER~\cite{10.1145/3544548.3580706} , SilentWhisper~\cite{10.1145/3706599.3721185}, DistillW2N~\cite{10888480} use self supervised learning to obtain whisper recognition capability based on HuBERT~\cite{Hsu2021-lz}. Recent speech recognition systems trained on large-scale vocabularies, such as OpenAI Whisper~\cite{radford2022robustspeechrecognitionlargescale}, are also capable of recognizing whispered speech. Farhadipour et al. also reported finetuning on OpenAI whisper on whisper utterances could increase recognition accuracy~\cite{farhadipour2024leveragingselfsupervisedmodelsautomatic}.

On the other hand, a significant challenge for whispered speech recognition is its high susceptibility to external noise.
WhisperMask addresses this with an input device designed to record the wearer's whispers while reducing external noise~\cite{10.1145/3652920.3652925}. Yet, because it necessitates wearing a mask, it is intended not for everyday use, but rather for environments where mask-wearing is already a prerequisite, such as medical facilities or construction sites.

Commercially available headsets, such as the Apple AirPods Pro, offer ``Voice Isolation'' feature that utilizes multi-microphone beamforming and signal processing to capture the wearer's speech~\cite{Apple2025VoiceIsolation}. However, our evaluation revealed that while this function works effectively for normal speech, it fails to capture whispered speech entirely (as demonstrated in the supplemental video). This suggests that distinguishing between whispered speech and external noise remains a significant challenge for standard signal processing algorithms.

Overall, many of these SSI studies do not address fully free-form speech but are instead designed for systems with limited vocabularies or for silent command input. In contrast, low-voice and whispered speech have reduced sound pressure, yet with speech transformation systems can maintain vocabulary size and recognition accuracy comparable to ordinary speech, allowing usage similar to SSI. However, since low-voice speech is more vulnerable to environmental noise, robust noise countermeasures are essential.

\subsection*{Voice Sensing with Bone/Skin Conduction}

Early NAM (non-audible murmur) work showed that skin-coupled stethoscopic/silicone microphones behind the ear can capture whispered or ``nonaudible murmur'' speech robustly to ambient noise, enabling ASR and even whisper-to-speech enhancement. Subsequent studies improved sensors and explored model adaptation and enhancement from body-conducted resonances. 

Moon investigated simultaneous bilateral recordings with mini-accelerometers on the nasal bones show that nasal airflow patency (resistance) affects the detectability of nasal-bone vibration~\cite{moon1990}. Measurement with an accelerometer over the nose is suitable for capturing nasal-origin energy but is modulated by nasal patency. 
Yiu et al. also examined the correlation between nasal-bridge bone vibration and perceptual auditory ratings~\cite{Yiu2012}. They find a moderate correlation between the magnitude of the vibration of the nasal bridge and the perceived resonance.

Chen et al. reported that resonant voice training increases facial, particularly nasal-bridge, bone vibration and suggested that these vibrations are likely to contribute to resonant voice production~\cite{CHEN2014596}. These findings support the usefulness of the nasal bridge as a measurement point. 
However, they did not examine the involvement of nasal-bridge vibration in whispered speech.

Kitamura used scanning laser Doppler vibrometry to map facial vibration velocity during speech~\cite{kitamura2012}. He reports strong vibrations at the nose and surrounding area for nasal sounds, and even for vowels, the lateral sides of the nose show prominent vibration. The area around the nose is visualized as a hot spot for speech-induced vibration.

V-Speech places a vibration sensor on the nose pads of smart glasses to capture speech~\cite{10.1145/3287058}. Compared with an air microphone, the SNR improves substantially making it practical for ASR and calls. However, ``nasal distortion'' can occur because nasal consonants become overamplified, so compensation must be provided.

In summary, while nasal sensing appears promising for speech acquisition, no prior work has attempted recognition by fusing bone/skin-conducted signals with conventional airborne speech.
However, whispered speech recognition using nasal sensing has not been explored in prior work.

\subsection*{Fused Air Conduction and Bone/Skin Conduction Utterance Recognition}



Zhou et al. use bone conduction (BC) for robust voice activation detection (VAD), adaptive filtering, and wind noise handling~\cite{Zhou2020_Sensors_BC_RealTime}. It demonstrates gains over an acoustic conduction (AC)-only array and describes a real-time embedded prototype.

Yu et al. uses a fully convolutional time domain architecture with early vs. late fusion of BC and AC~\cite{Yu2020_SPL_BoneAir_TimeDomain}. It reports that late fusion performs better and surpasses single-modality enhancement on a Mandarin corpus.
Wang et al. propose an attention-based fusion of BC and noisy AC spectrograms~\cite{Wang2022_TASLP_ACBCFusion}. It also introduces a semi-supervised training scheme (CycleGAN-style) to leverage unpaired AC/BC data. On the EMSB corpus it outperforms a time-domain baseline and single-sensor systems, with especially strong gains at low SNRs. 
Wang et al. introduce an end-to-end time-domain model (MMINet) that jointly uses noisy AC + BC to enhance speech under low SNR conditions, yielding consistent improvements over AC-only and BC-only models~\cite{Wang2022_AppliedAcoustics_MMINet}.

VibVoice targets wearables by fusing microphone audio + on-device IMU sensors~\cite{He2023_MobiSys_VibVoice}. It shows BC vibrations are insensitive to external speakers and mostly below ~800 Hz, then proposes a Bone Conduction Function for data augmentation and trains a multi-modal DNN that improves enhancement under competing-speaker and motion conditions. However, whispered voice input is not achieved by this work.
AccCall also utilizes an audio microphone and the IMU built into a smartphone to reduce speech noise~\cite{10.1145/3749463}. However, this study has also not realized the recording of whispered speech.

Huang et al. highlight that conventional BC + AC fusion can degrade sharply with even tiny mismatches~\cite{Huang2024_SignalProcessing_OnlineFusion}. They propose an online adaptive fusion method suitable for real-time communication scenarios. Useful for deployment where sensor latencies/placements drift. 
Kuang et al. build a DenGCAN backbone with attention-based feature fusion and attention-gated skip connections~\cite{Kuang2024_JASA_LightweightFusion}. They report improvement over noisy AC, showing that compact BC+AC fusion models can still be highly effective.

\section{NasoVoce}

\begin{figure}
\begin{center}
\includegraphics[width=0.95\columnwidth]{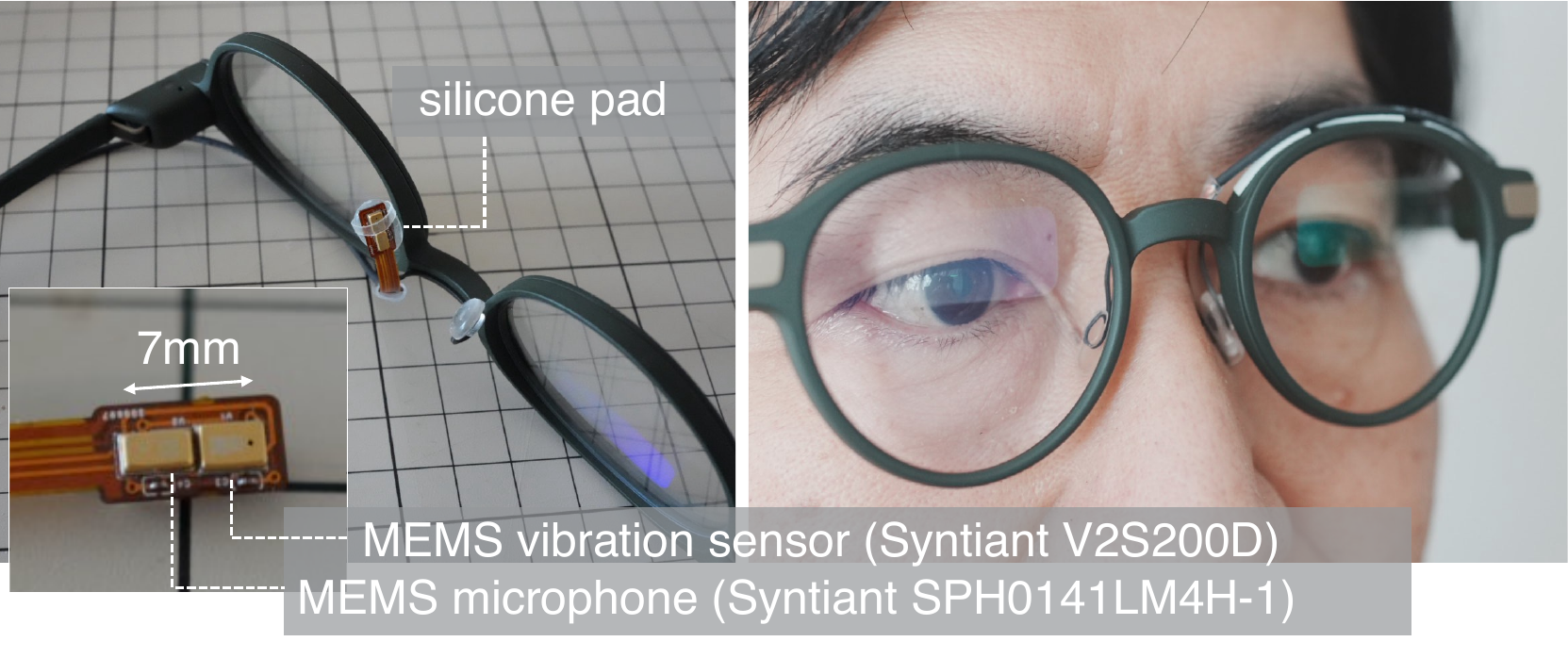}
\end{center}
\caption{NasoVoce sensor configuration: A MEMS microphone and MEMS vibration sensor can be acquired as time-synchronized digital data as the left and right channels of a TDM audio interface. An example usage is shown with the sensor mounted on the nose pad of a smart glasses frame.}
\label{fig:config}
\end{figure}

The NasoVoce sensor configuration is illustrated in Fig.~\ref{fig:config}. It is composed of a MEMS microphone (Syntiant SPH0141LM4H-1~\cite{SyntiantMEMS}) and a MEMS vibration sensor (Syntiant V2S200D~\cite{V2S200d}). Both devices provide a pulse density modulation (PDM) interface with 16KHz sampling rate and 16-bit digitization, allowing them to be driven by the same clock. This configuration enables the acquisition of accurately time-synchronized acoustic signals from both sensors at the clock in the digital domain, eliminating the time-mismatch problem pointed out by Huang et al.~\cite{Huang2024_SignalProcessing_OnlineFusion}.

By mapping the vibration sensor signal to the left channel and the microphone signal to the right channel, the sensor output can be treated as a conventional stereo audio stream, thus maintaining compatibility with a wide range of existing audio interfaces. The V2S200D vibration sensor is capable of measuring at higher frequencies (up to 10 kHz) than typical IMU (Inertial Measurement Unit) sensors, making it well suited for capturing skin- or bone-conducted speech signals including whisper utterances.

Fig.~\ref{fig:config} (right) shows an example in which the sensor is mounted on the nose pad of a commercially available smart glasses frame. The design incorporates a silicone pad that mechanically isolates vibrations from the frame while ensuring a discreet appearance suitable for continuous wear.

\begin{figure}
\begin{center}
\includegraphics[width=0.8\columnwidth]{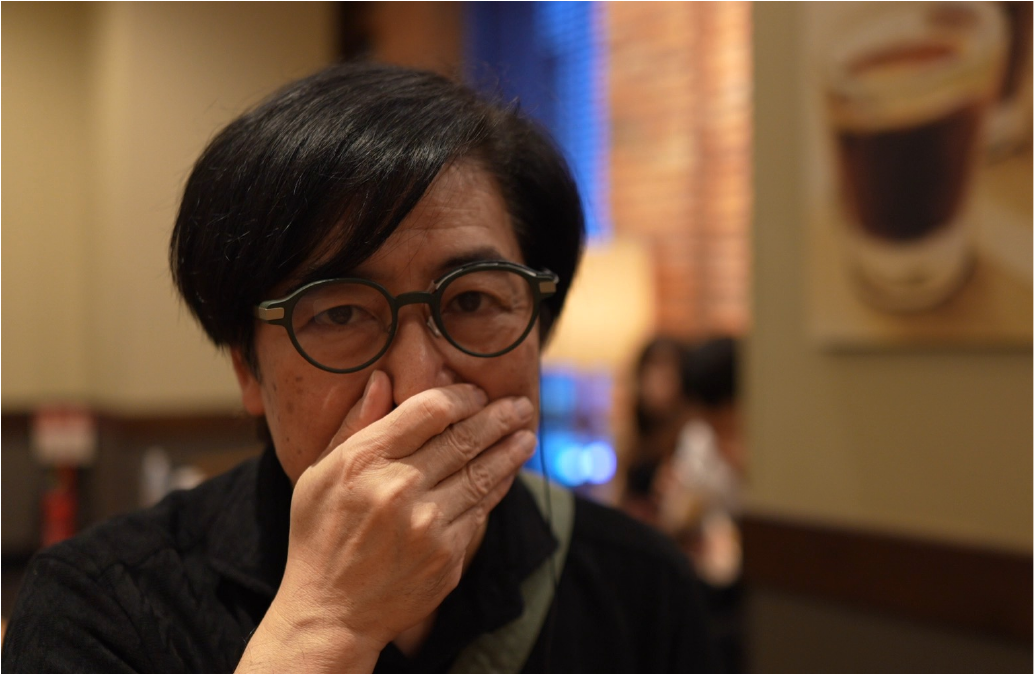}
\end{center}
\caption{By covering the mouth and nose with one’s hand, the spoken content cannot be inferred through lip reading. This posture further serves as a social signal, indicating that the person is engaged in a conversation with the device.}
\label{fig:hand}
\end{figure}

This configuration provides two key advantages: (1) speech can be captured via bone/skin conduction from the nasal bones, and (2) proximity to the mouth makes it effective for capturing air-conducted speech. Although whispered speech tends to propagate poorly through bone conduction, we will show it can still be acquired through the air-conduction channel of the microphone. For private speech input, users can cover their mouth with a hand, which prevents sound leakage into the environment (Fig.~\ref{fig:hand}). This gesture not only protects the secrecy of speech, but also functions as a socially interpretable cue, signaling to bystanders that the user is engaging in voice-based interaction with a device.

\section{Recognition Model}

\begin{figure*}
\begin{center}
\includegraphics[width=0.8\textwidth]{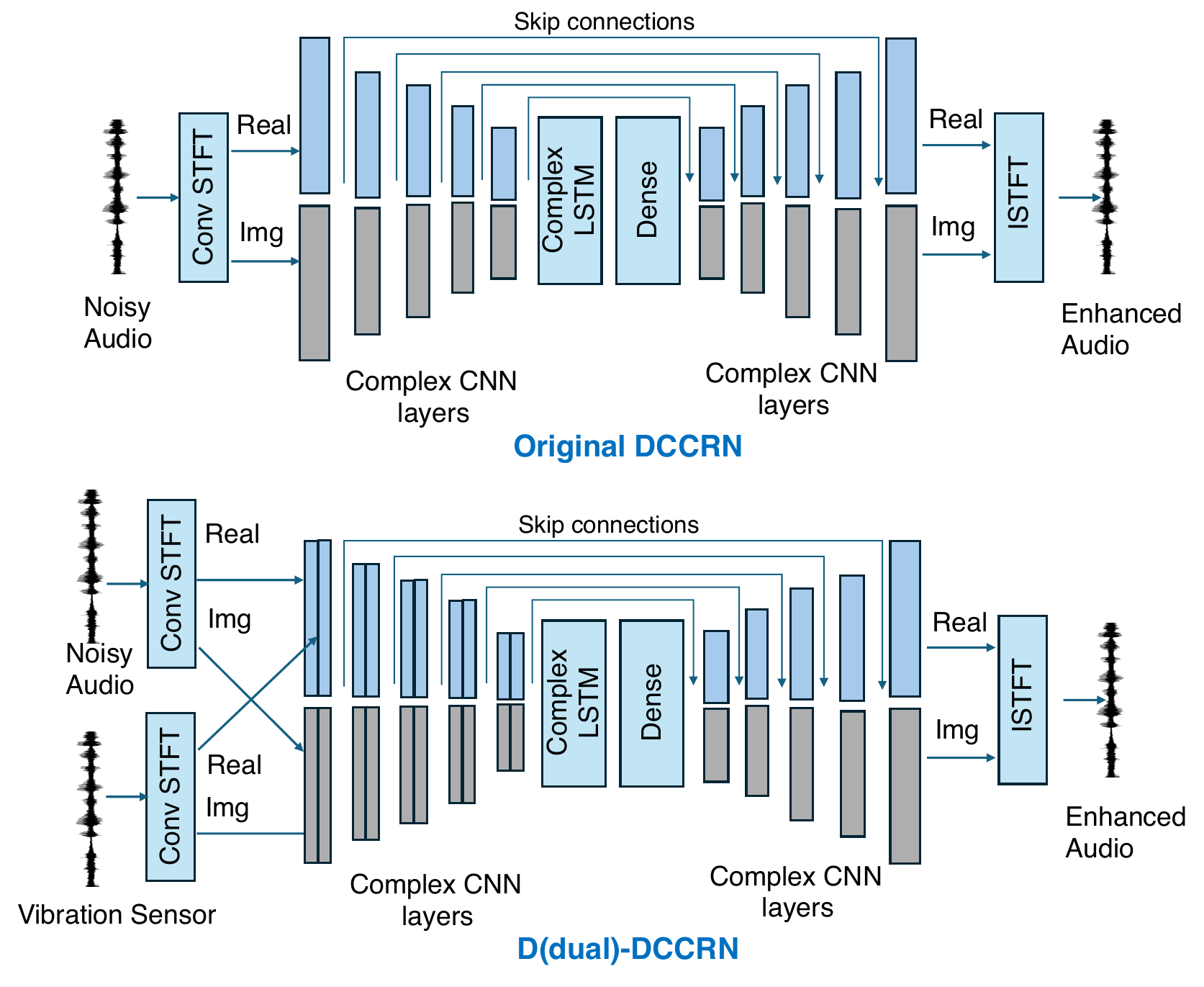}
\end{center}
\caption{D-DCCRN (Dual-DCCRN) accepts composite inputs from a microphone (Mic) and a vibration sensor (Vib). D-DCCRN generalizes the design of DCCRN audio enhancement model to jointly process the real and imaginary components of both Mic and  Vib signals.}
\label{fig:DCCRN}
\end{figure*}

\begin{figure}
\begin{center}
\includegraphics[width=0.95\columnwidth]{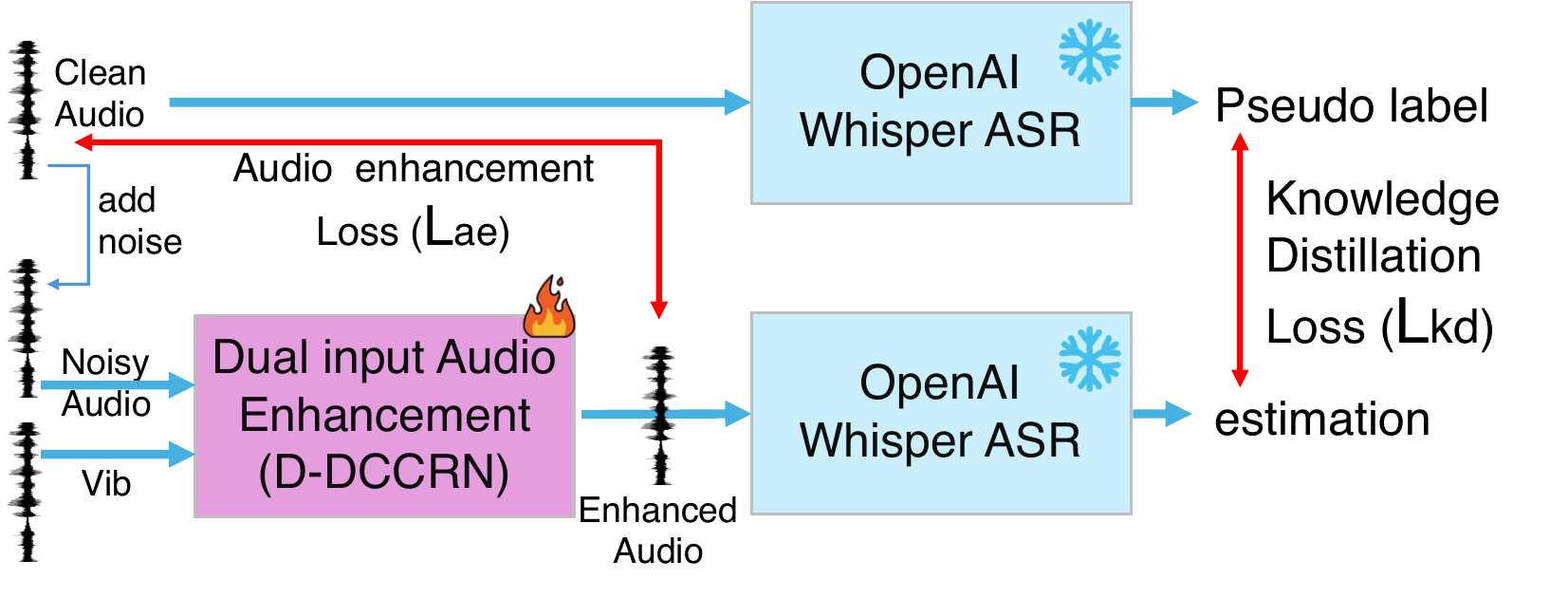}
\end{center}
\caption{The NasoVoce training method: it combines audio enhancement loss ($L_{ae}$) and Knowledge Distillation loss ($L_{kd}$)}.
\label{fig:train}
\end{figure}

We constructed and evaluated a neural network recognition model that combines the vibration sensor (Vib) and microphone (Mic) input : D-DCCRN by extending DCCRN (Fig.~\ref{fig:DCCRN}).

DCCRN (Deep Complex Convolution Recurrent Network)~\cite{DCCRNinterspeech20} is a deep learning model for audio enhancement that combines a complex-valued convolutional encoder–decoder with a complex-valued LSTM (Long Short-Term Memory), allowing effective handling of both magnitude and phase. 
We adopted the DCCRN architecture because whispered speech lacks harmonic structure and closely resembles noise, making it difficult for conventional magnitude-based networks to separate from environmental noise. DCCRN utilizes complex-valued networks to process both magnitude and phase information. Since the ``phase'' structure of the near-field whisper turbulence (captured by the vibration sensor) differs significantly from far-field environmental noise, we expect preserving phase information is crucial for effective enhancement.

The extended model is named D-DCCRN (Dual-DCCRN); it accepts composite inputs from a conventional microphone (Mic) and a vibration sensor (Vib). While the original DCCRN applies a short-time Fourier transform (STFT) and processes the real and imaginary parts using complex CNN layers, D-DCCRN generalizes this design to jointly process the real and imaginary components of both Mic and Vib signals.
Same as DCCRN, the output of D-DCCRN is enhanced speech, enabling direct listening, communication use, or input to any ASR system. The main advantage is improved audio quality by leveraging both Vib and Mic features. 

D-DCCRN has 438.25M parameters, which is \qty{28}{\%} of the parameter size of OpenAI Whisper large-v2 (\qty{1543.3}{M}).
The average processing time for D-DCCRN is \qty{136.9}{ms}, while the average processing time for OpenAI Whisper is \qty{429.96}{ms}, which means that D-DCCRN requires \qty{31.8}{\%} of Whisper’s processing time.
We consider this computational cost acceptable in light of the benefit of noise-robust whispered speech recognition. However, fully streaming speech processing and integration into smartphones have not yet been realized and remain as future work.


\subsection{Training Method}

We collected a data set of clean speech and vibration sensor paired signals. 45 participants (English fluent, gender balanced, 25--55 years old) each read English text obtained from the Free ST American English Corpus~\cite{freeSTAmerican} for approximately 2.3 hours, recorded simultaneously with a MEMS microphone and a vibration sensor, resulting in a total of 104 hours. To simulate noisy environments, clean speech audio signals were mixed with samples from the DEMAND noise dataset~\cite{DEMAND}. Noise was added at root mean square (RMS) levels ranging from \qty{-10}{dB} to \qty{+10}{dB} relative to the clean speech RMS level, with noise instances randomly selected from the noise dataset for each utterance to ensure diverse corruption patterns.

\begin{figure*}
  \centering
  \includegraphics[width=0.85\textwidth]{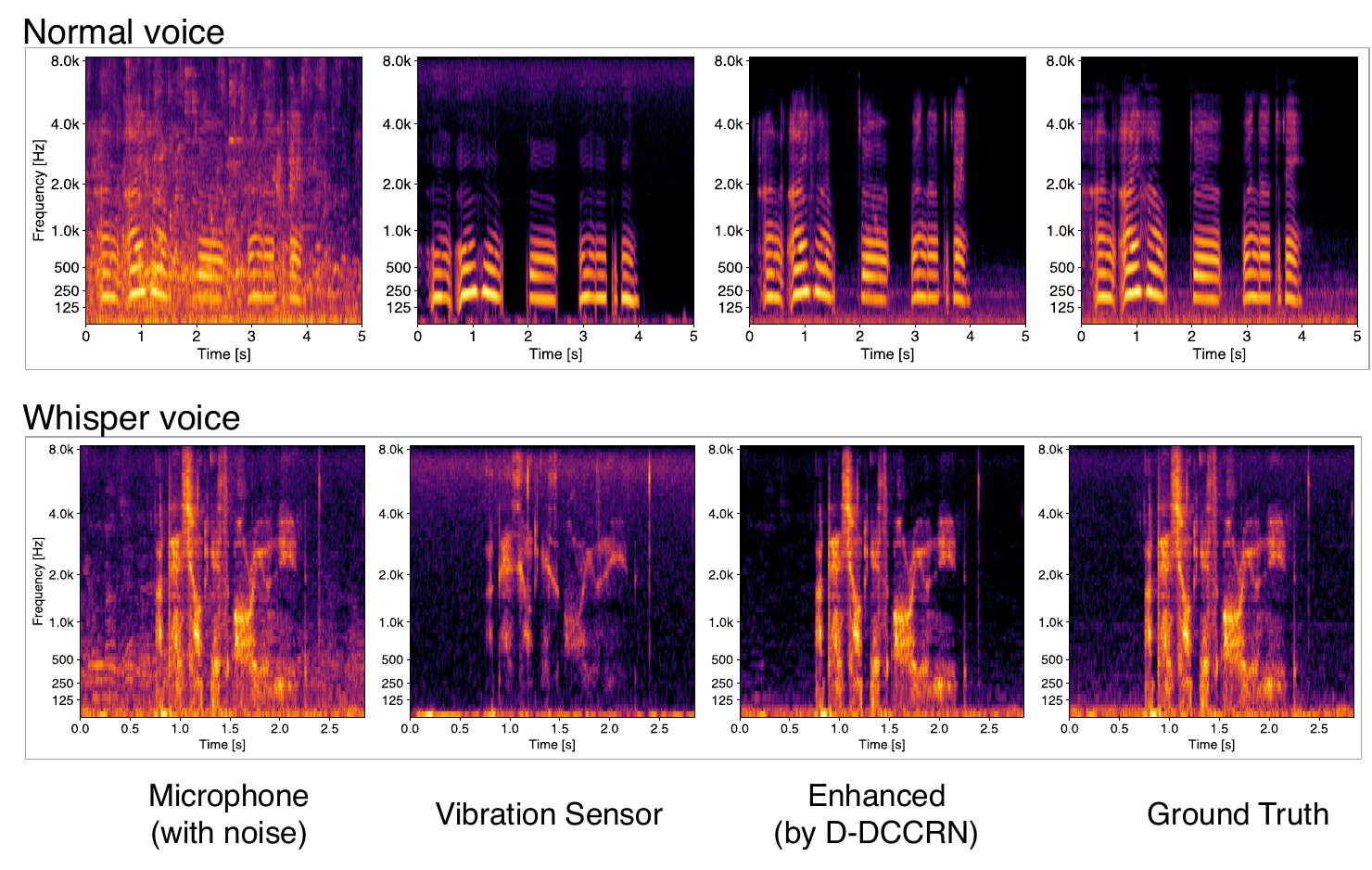}
  \caption{Audio quality improvement examples (normal and whispered speeches) through the combined use of microphone (Mic) input and vibration sensor (Vib) input:
  While the Mic input contains external noise, the audio enhancement results obtained from both Mic and Vib inputs (Enhancement) closely approximate the Ground Truth. This outcome is demonstrated through a simulation in which noise is added to the Ground Truth to form the Mic input, and an audio enhancement model is subsequently applied.
}
  \label{fig:mel}
\end{figure*}

For D-DCCRN, the objective was to reconstruct clean speech from Vib $+$ noisy audio input (Mic). We trained the model using a reconstruction loss targeting clean audio. In addition, we introduced a loss based on the OpenAI Whisper decoder, inspired by knowledge distillation methods~\cite{journals/corr/HintonVD15} such as Distil-Whisper~\cite{gandhi2023distilwhisperrobustknowledgedistillation}. The recognition results from the Vib $+$ noisy input are trained to match those from the clean input. We applied both a soft loss (aligning decoder output distributions) and a hard loss (improving token-level accuracy) following standard distillation practice. Fig.~\ref{fig:mel} shows an example of audio enhancement from noisy audio and vibration sensor.

At the audio enhancement level, loss is defined as:

\begin{equation}
\begin{aligned}
enc &= {AE}_{dual}(Noise(mic), vib) \\
L_{ae} &= MSE(Mel(mic), Mel(enc)) + \lambda_{si} Si\text{-}SDR(mic, enc)
\end{aligned}
\end{equation}

\noindent
where $mic$ is mic signal, $vib$ is vibration signal, $AE_{dual}(\cdot,\cdot)$ is dual input audio enhancement (D-DCCRN), $Noise(\cdot)$ is a noise adding function, $MSE(\cdot)$ is mean square error, $Mel(\cdot)$ is mel-spectrogram, $SI\text{-}SDR$ is Scale-Invariant Signal-to-Distortion Ratio, and the goal is to minimize the difference between the clean encoder output ($enc$) and the noise-robust encoder output ($enc'$).

To ensure consistency at the ASR level, we additionally use $L_{kd}$, loss of knowledge distillation following the method proposed by Distil-Whisper~\cite{gandhi2023distilwhisperrobustknowledgedistillation}. In Distill-Whisper, the output of the OpenAI Whisper (decoder) is used as a teacher signal to enhance the capability of the student decoder. In contrast, our objective is to improve the ASR performance of the dual input audio enhancement output. To this end, we freeze both the encoder and decoder of OpenAI Whisper and use them solely as an evaluator, while employing them in the training process of the audio enhancement model.

To achieve this objective, we define two losses: a hard loss $L_{hard}$ and a soft loss $L_{soft}$~\cite{journals/corr/HintonVD15}.
Using the OpenAI Whisper encoder, we get hidden state representation $H$ from clean $Mic$ input and hidden state representation $H'$ from noise enhanced input $enc$. Then $L_{hard}$ is defined as:

\begin{equation}
L_{hard} = - \displaystyle\sum_{i=1}^{N'} P(y_i | \hat{y} < i, H')
\end{equation}

\noindent
where $\hat{y}$ are the pseudo-labels~\cite{kim2016sequencelevelknowledgedistillation} generated from clean speech. This loss represents how accurately enhanced audio from noisy voices can generate tokens compared to clean voice tokens.

$L_{soft}$ can be defined as the Kullback–Leibler (KL) divergence of the probability distributions of clean audio ($Q_i$) and enhanced audio ($R_i$), so that $R_i$ can be trained to match the full distribution of the teacher ($Q_i$) by minimizing the KL divergence over the entire set of token probabilities at position $i$:

\begin{equation}
L_{soft} = \displaystyle\sum_{i=1}^{N} KL(Q_{i}, P_{i})
\end{equation}

\noindent
Then we get the combined knowledge distillation (KD) objective:
 \begin{equation}
 L_{kd} =\lambda_{soft}L_{soft} +\lambda_{hard}L_{hard}
 \end{equation}

\noindent
Finally, the total loss is given by:

\begin{equation}
L_{total} = L_{ae} + \lambda L_{kd}
\end{equation}

\section{Evaluation}

\begin{figure}
\begin{center}
\includegraphics[width=1.0\columnwidth]{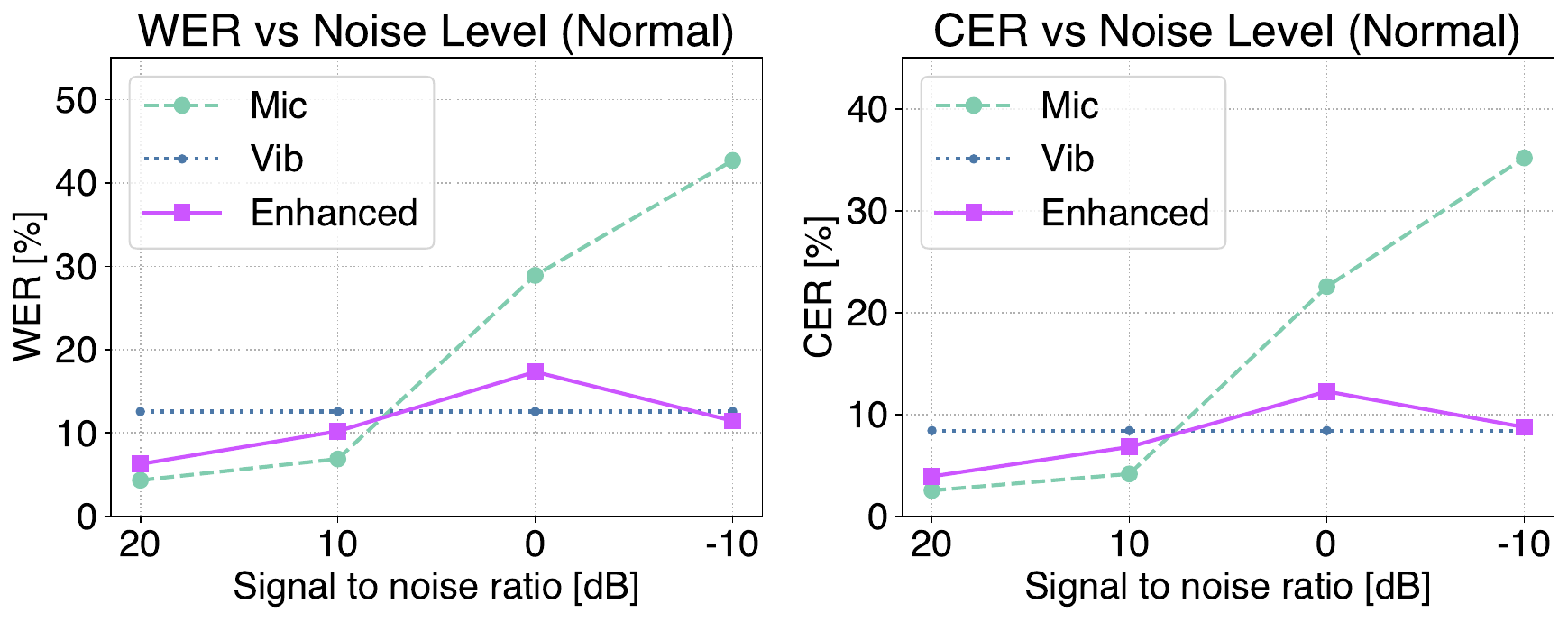}
\includegraphics[width=1.0\columnwidth]{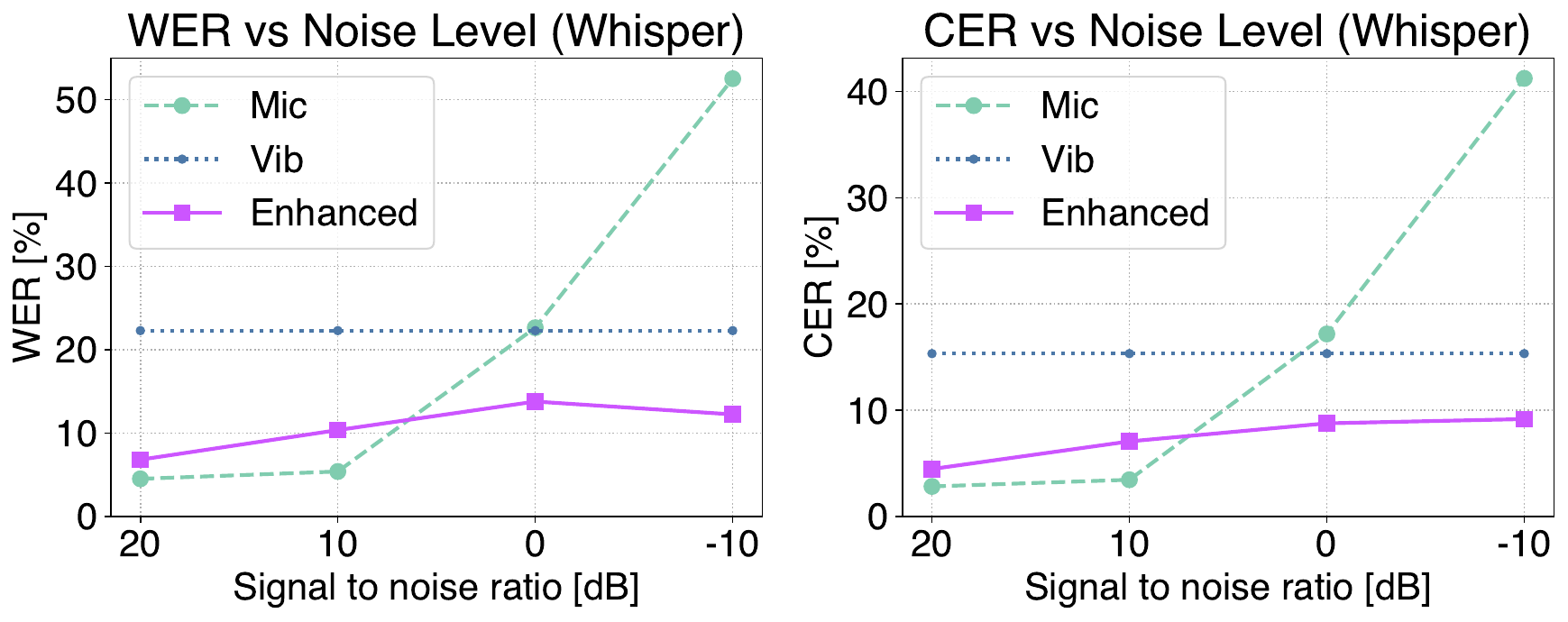}
\end{center}
\caption{Speech recognition accuracy (WER, CER) for normal and whispered speech:  MEMS microphone (Mic), MEMS vibration sensor (Vib), audio enhancement by D-DCCRN}
\label{fig:wer}
\end{figure}

\begin{figure}
\begin{center}
\includegraphics[width=1.0\columnwidth]{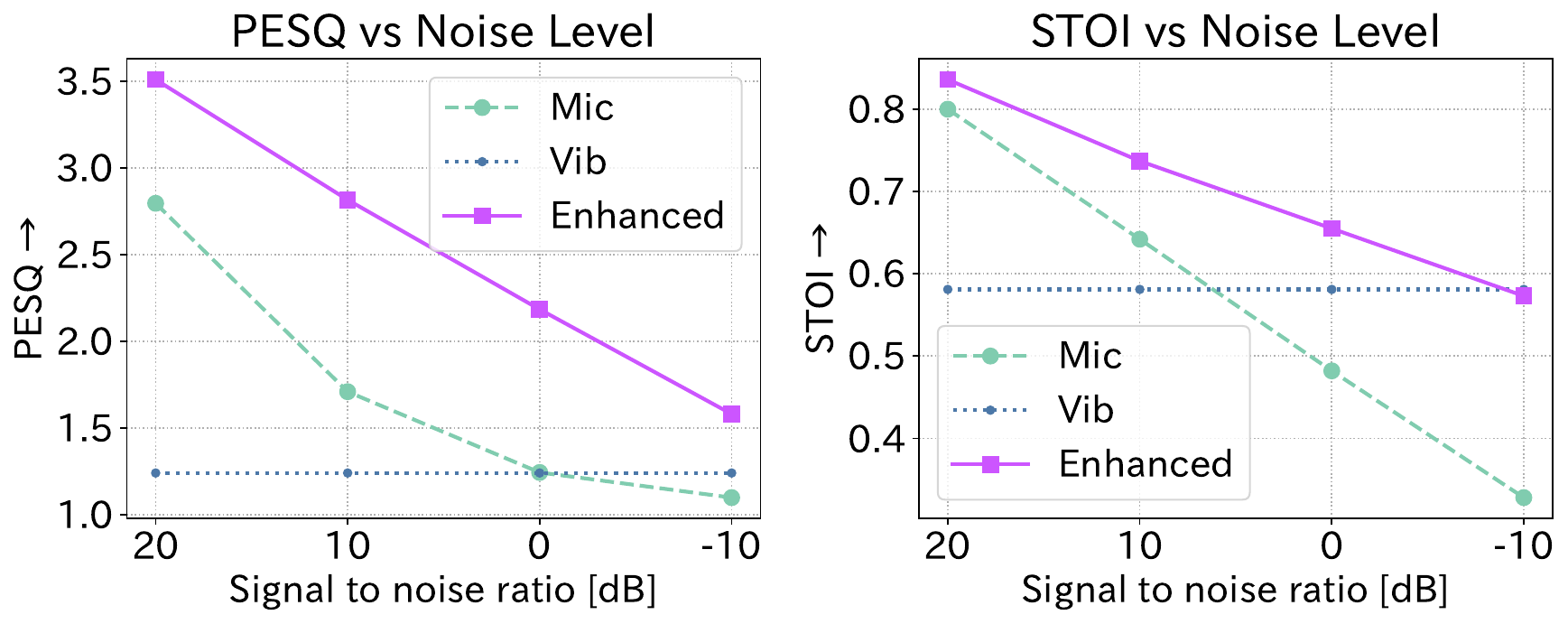}
\end{center}
\caption{PESQ and STOI scores of MEMS microphone (Mic), MEMS vibration sensor (Vib), and enhanced voice using the D-DCCRN model by combining Mic and Vib (D-DCCRN) under various noise conditions.}
\label{fig:pesq}
\end{figure}

\begin{figure}
\begin{center}
\includegraphics[width=0.95\columnwidth]{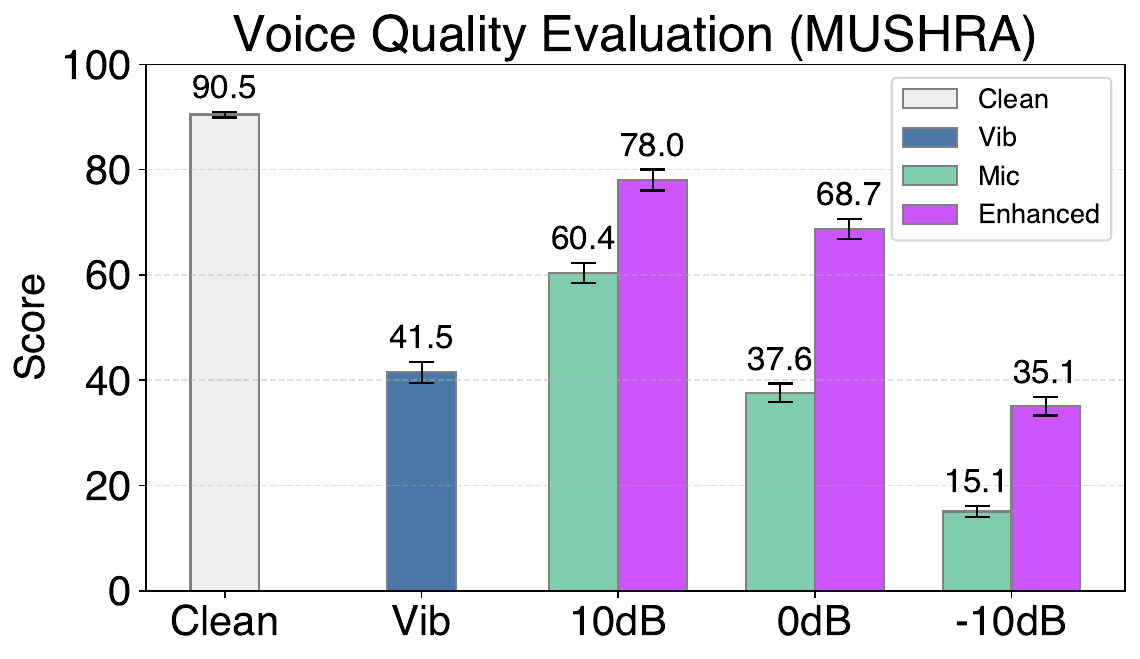}
\end{center}
\caption{MUSHRA-based audio quality evaluation: We assessed the Clean speech (Hidden Reference), the simultaneously recorded MEMS vibration sensor (Vib), a microphone signal with added noise (Mic, SNR \qty{10}{dB}, \qty{0}{dB}, \qty{-10}{dB}), and an Enhanced signal obtained by fusing Mic and Vib and processing with D-DCCRN.}
\label{fig:mushra}
\end{figure}

To evaluate the proposed approach, we conducted (i) ASR accuracy tests, (ii) objective audio-quality measurements, and (iii) participant-based quality ratings. The evaluation data were disjoint from the training set; for each utterance we prepared a clean source and a corresponding noise-corrupted version. ASR and objective quality were computed on 1,000 items. For the subjective study, participants rated five sets, each comprising: a clean reference, a simultaneously recorded MEMS vibration sensor (Vib) channel, microphone signals obtained by mixing the clean speech with various noises (Mic conditions), and signals enhanced from Vib+Mic by the proposed audio-enhancement model.

\subsection{ASR (WER /CER)}

Fig.~\ref{fig:wer} reports ASR recognition accuracy in terms of Word Error Rate (WER) and character error rate (CER). 
We conducted the measurements using data consisting of 200 utterances from each of the evaluators (n = 4) who were distinct from the participants involved in the training dataset collection. As in the training phase, noise was superimposed using the DEMAND noise dataset~\cite{DEMAND}.
Mic denotes recognition from the microphone audio; Vib denotes recognition directly from the vibration sensor signal; Enhanced denotes recognition from an Enhanced signal produced by audio enhancement that fuses Mic and  Vib.

As shown in the figure, for both normal and whispered speech, Mic recognition accuracy degrades as the noise level increases. While Vib remains relatively stable, its recognition accuracy for whispered speech is worse than that of Normal. Enhanced is comparatively robust to noise and, in particular, consistently achieves better recognition accuracy than Vib for whispered speech. Moreover, when the noise level is \qty{0}{dB} or higher, Enhanced outperforms Mic in terms of recognition accuracy.
From these results, we can conclude that NasoVoce can reliably recognize whispered speech even under noisy conditions. 

On the other hand, for normal speech, depending on the situation, Vib can sometimes achieve the highest recognition accuracy. In previous mixed-modality speech recognition studies, the IMU sensing bandwidth was around 0–400 Hz, which is not directly suitable for speech recognition. In contrast, our system employs a MEMS vibration sensor (V2S200D~\cite{V2S200d}) with a sensing range much higher than normal IMUs (10Hz - 2000Hz), suggesting that Vib alone may be sufficient for speech recognition. As future work, we plan to develop a recognition model that can dynamically select the input configuration -- Mic, Mic+Vib, or Vib alone -- that yields the highest recognition accuracy under a given condition.

\subsection{Voice Quality (PESQ / STOI)}

We then evaluated voice enhancement quality with
PESQ (Perceptual Evaluation of Speech Quality)~\cite{ITU-T-P862-2001,Rix2001PESQ} and STOI (Short-Time Objective Intelligibility)~\cite{Taal2010STOI,Taal2011STOI}.
PESQ is an intrusive reference-based metric (ITU-T P.862/P.862.2) that predicts perceived quality by comparing a processed/degraded signal to a clean reference using a psychoacoustic model.
STOI is an intrusive metric that estimates speech intelligibility by correlating short-time temporal envelopes across one-third octave bands between clean and processed speech.

In our case, MEMS microphone (Mic), MEMS vibration sensor (Vib), and enhanced voice using the D-DCCRN model by combining Mic and Vib (Enhanced), are evaluated (Fig.~\ref{fig:pesq}), under various noise conditions (\qty{-20}{dB}, \qty{-10}{dB}, \qty{0}{dB}, and \qty{10}{dB}). As shown in the graph, the enhanced results outperform Mic conditions in all noise levels. Under very strong noise (\qty{10}{dB}), the Vib condition is better than the Enhanced model.

\subsection{User Evaluation (MUSHRA)}

Fig.~\ref{fig:mushra} shows the subjective evaluation of speech quality with MUSHRA (MUltiple Stimuli with Anchor)~\cite{mushra}. MUSHRA is a method to evaluate audio quality as defined by ITU-R Recommendation BS.1534. It uses hidden reference speech and other speech, and participants are expected to give scores (from 0 to 100) comparing the anchor audio as a reference. 
Due to the use of a hidden reference, MUSHRA is considered to be more reliable than the mean opinion score (MOS).
In our case, a clean voice was used as a reference. The participants rated the Clean voice (Hidden Reference), Vib (audio from the MEMS vibration sensor), Mic (audio from the MEMS microphone), and Enhanced (enhanced audio by the proposed D-DCCRN model with Vib and Mic input) under various noise conditions (\qty{-10}{dB}, \qty{0}{dB}, and \qty{10}{dB}). 

50 English-fluent, gender-balanced participants over 18 years of age were recruited using the Prolific online survey system~\cite{prolific} with a MUSHRA online tool based on JavaScript~\cite{mushraJS}. For each rating, participants could play each voice as many times as they wanted.

As shown in the graph, the Enhanced condition consistently outperformed the Mic condition and remained superior to Vib until the noise level reached \qty{0}{dB}. Under the very high noise condition (\qty{10}{dB}), the Vib was slightly higher. We hypothesize that in high-noise environments, the microphone input contributes minimally to the Enhanced model's output, thereby limiting improvements in perceived quality.

\begin{figure*}
\begin{center}
\includegraphics[width=0.95\textwidth]{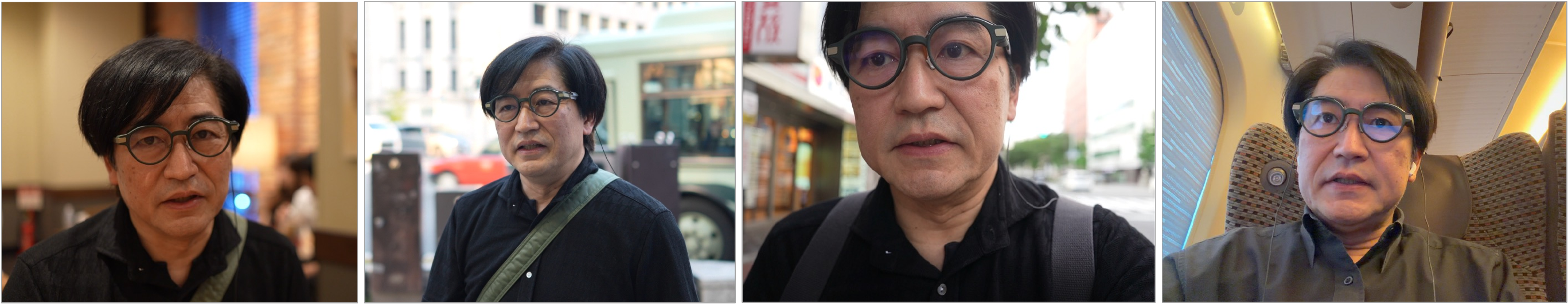}
\end{center}
\caption{In-the-wild trial of NasoVoce: We recorded normal and whispered speech in four everyday environments—inside a café, on a busy roadside, while walking, and inside a train car—and compared signals obtained with a conventional microphone and with NasoVoce.
}
\label{fig:inthewild}
\end{figure*}

\subsection{In-the-Wild Test}

The evaluations described above were conducted with artificially superimposed noise. We also evaluated the acoustic performance of NasoVoce in real-world environments. Specifically, we recorded both normal and whispered speech in four scenarios: inside a café, on a busy roadside, while walking outdoors, and inside a train car. 

For each setting, we captured speech simultaneously with a conventional air-conduction microphone and with NasoVoce, and compared the resulting signals. Fig.~\ref{fig:inthewild} illustrates the experimental setup; representative audio is included in the supplementary video.

In the in-train condition, we additionally compared NasoVoce against the ``voice isolation feature'' of Apple AirPods Pro 2. The voice isolation function performs multi-microphone beamforming and iOS-side signal processing to suppress background sounds other than the wearer’s speech. The comparison showed that, while Voice Isolation effectively attenuated environmental noise for normal speech, it almost completely suppressed whispered speech, presumably because whisper components were treated as background noise. In contrast, NasoVoce consistently captured whispered speech across all environments while robustly attenuating external noise.

Overall, our results show that the proposed dual sensor method (Vibration + Mic) achieves higher recognition accuracy than Mic alone under diverse noise conditions and remains superior to vibration alone until the noise level reaches \qty{0}{dB} (i.e., noise and speech at parity). This is supported by recognition accuracy on the trial user's utterances (WER, CER), objective quality / intelligence metrics (PESQ, STOI), and user ratings using the MUSHRA procedure.

\section{Discussions}

\subsection*{Use of Whisper, Vibration, and Enhanced Speech}

According to our evaluation results, under low-noise conditions, the accuracy of Enhanced speech and Mic speech is comparable. However, as noise levels increase, the accuracy of Mic speech drops sharply. In contrast, under very high-noise conditions, the performance of Enhanced speech becomes roughly equivalent to that of the Vib input alone.

For whispered speech, however, the performance of Enhanced speech degrades considerably under noisy conditions. We attribute this to the small vibration signals produced during whispering, which reduce the quality of the Vib input. On the other hand, in quiet environments, whisper input is particularly valuable, while in environments with substantial external noise, whisper input is less necessary, and normal speech can be used instead.

We argue that it is desirable to provide either a user interface that allows flexible selection of the optimal input mode or an automatic mechanism that determines the most suitable input method depending on context.

\subsection*{Adaptive Sensor Fusion Strategy}

While our dual-input model (D-DCCRN) achieves good performance across most conditions, we observed that under extreme noise conditions (e.g., \qty{+10}{dB}), the recognition accuracy using the vibration sensor (Vib) alone exceeds that of the enhanced signal. This indicates that when the acoustic Signal-to-Noise Ratio (SNR) drops below a critical threshold, the air-conducted microphone signal may introduce more noise artifacts than useful speech features into the fusion network, counteracting the benefits of multi-modal integration.

This finding provides a design implication for real-world deployment: a robust always-available speech interface should not rely on a static fusion weight. Instead, we could propose an \textit{SNR-adaptive gating mechanism}. Such a system would monitor environmental noise levels in real-time and dynamically transition between modes.
The ability to discard the microphone signal entirely in favor of the mechanically isolated vibration sensor validates the necessity of our heterogeneous sensor configuration, ensuring communication remains possible even when air-conduction fails.

\subsection*{Alternative Sensor Configurations}

\begin{figure*}
\begin{center}
\includegraphics[width=0.8\textwidth]{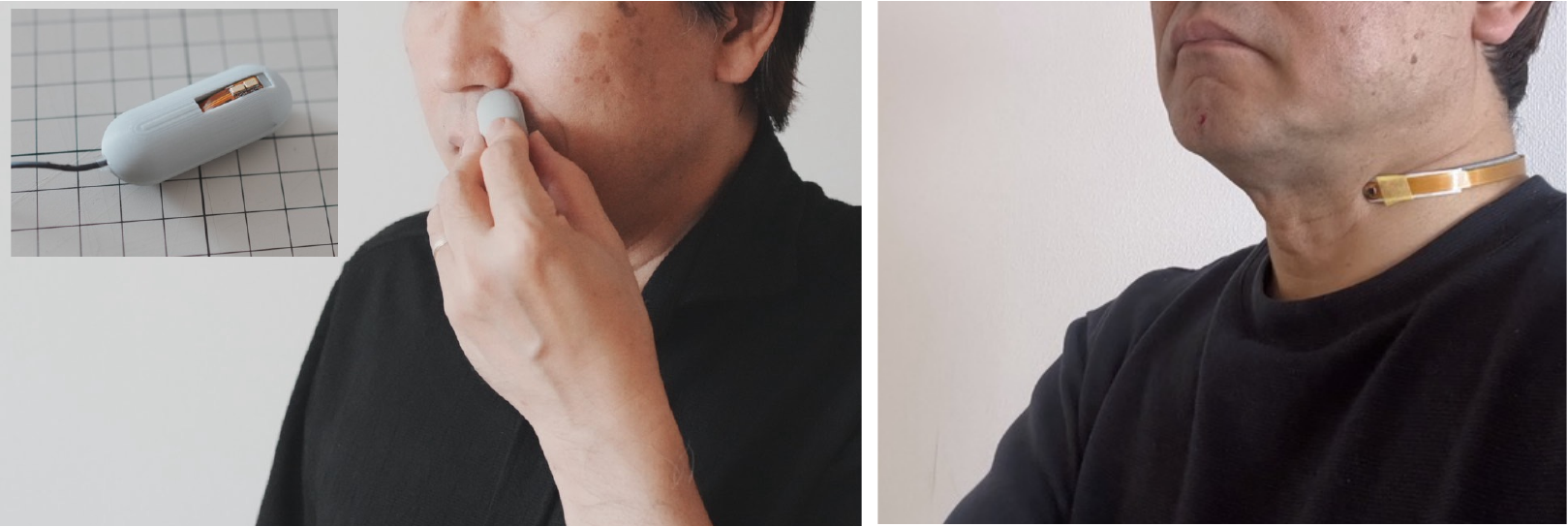}
\end{center}
\caption{Alternative use cases of the proposed Vib-Mic sensor:
(Left) Placed against the skin between the upper lip and the nose, it can reliably capture whispered speech via skin conduction.
(Right) Applied as a throat microphone, it acquires both skin-conducted and vocal signals through the neck.}
\label{fig:alt}
\end{figure*}

We expect the proposed dual-sensor approach, which combines a vibration sensor (Vib) and a microphone (Mic), to remain effective at contact sites other than the nose. Fig.~\ref{fig:alt} (left) illustrates use with the device placed against the philtrum (the region between the upper lip and the nose). In a non-wearable, voice-memo–style configuration, the device can operate as a conventional microphone during normal phonation and be pressed against the skin in noisy environments or when whispered speech is required. Fig.~\ref{fig:alt} (right) also depicts a throat-microphone configuration in which the Vib and Mic are co-located; compared with throat microphones that rely solely on skin conduction, this design is expected to improve audio quality.

\subsection*{Physiological Variability}
Previous work suggests that nasal patency modulates accelerometric detectability on the nasal bridge. Given our nose‑pad mounting, future versions should incorporate
per-user calibration and online adaptation to mitigate day‑to‑day variability (e.g., rhinitis, mask wear).

\section{Conclusion}

We present NasoVoce, a nose bridge-mounted interface that integrates a microphone and a vibration sensor. Positioned at the nasal pads of smart glasses, NasoVoce unobtrusively captures normal and whispered utterances in everyday use, through acoustic and vibration sensors. 

The nasal bridge's proximity to the mouth facilitates the acquisition of bone- and skin-conducted speech. Compared with earphone or neckband devices, it can also capture low-volume utterances, such as whispered speech, more reliably.
While the microphone provides high-quality audio, it is susceptible to environmental noise; whereas the vibration sensor captures speech through bone and skin conduction with strong noise robustness, but lower quality. By combining these complementary inputs, we obtain high-quality speech that is robust to external noise.
We evaluated two dual-input models: D-DCCRN, an extension of the DCCRN audio enhancement model.
Recognition experiments with OpenAI Whisper Large-v2 ASR confirm that the proposed models improve accuracy under noisy conditions. Objective metrics (PESQ, STOI) and subjective ratings (MUSHRA) further validate the benefits of multimodal input.
This work demonstrates the feasibility of NasoVoce as a practical interface for continuous and discreet AI voice conversations.

\begin{acks}
This work was supported by JST Moonshot R\&D Grant JPMJMS2012, 
JSPS KAKENHI Grant Numbers JP25097211, JP24984259.
We thank Yoko Rekimoto for her assistance with the demonstration video production.
\end{acks}

\bibliographystyle{ACM-Reference-Format}
\bibliography{reference}

@online{Apple2025VoiceIsolation,
  author  = {{Apple Inc.}},
  title   = {Use Voice Isolation, Wide Spectrum, or Automatic Mic Mode on your iPhone and iPad},
  year    = {2025},
  url     = {https://support.apple.com/en-us/101993},
  note    = {Accessed: 2025-11-29}
}

@article{CHEN2014596,
title = {Facial Bone Vibration In Resonant Voice Production},
journal = {Journal of Voice},
volume = {28},
number = {5},
pages = {596-602},
year = {2014},
issn = {0892-1997},
doi = {https://doi.org/10.1016/j.jvoice.2013.12.014},
url = {https://www.sciencedirect.com/science/article/pii/S0892199713002786},
author = {Fei C. Chen and Estella P.-M. Ma and Edwin M.-L. Yiu}
}

@INPROCEEDINGS{10888480,
  author={Tan, Tianyi and Ruan, Haoxin and Chen, Xinan and Chen, Kai and Lin, Zhibin and Lu, Jing},
  booktitle={ICASSP 2025 - 2025 IEEE International Conference on Acoustics, Speech and Signal Processing (ICASSP)}, 
  title={DistillW2N: A Lightweight One-Shot Whisper to Normal Voice Conversion Model Using Distillation of Self-Supervised Features}, 
  year={2025},
  volume={},
  number={},
  pages={1-5},
  doi={10.1109/ICASSP49660.2025.10888480}}

@inproceedings{10.1145/3652920.3652925,
author = {Hiraki, Hirotaka and Kanazawa, Shusuke and Miura, Takahiro and Yoshida, Manabu and Mochimaru, Masaaki and Rekimoto, Jun},
title = {WhisperMask: a noise suppressive mask-type microphone for whisper speech},
year = {2024},
isbn = {9798400709807},
publisher = {Association for Computing Machinery},
address = {New York, NY, USA},
url = {https://doi.org/10.1145/3652920.3652925},
doi = {10.1145/3652920.3652925},
booktitle = {Proceedings of the Augmented Humans International Conference 2024},
pages = {1–14},
numpages = {14},
location = {Melbourne, VIC, Australia},
series = {AHs '24}
}

@inproceedings{10.1145/3666025.3699374,
author = {Srivastava, Tanmay and Khanna, Prerna and Pan, Shijia and Nguyen, Phuc and Jain, Shubham},
title = {Unvoiced: Designing an LLM-assisted Unvoiced User Interface using Earables},
year = {2024},
isbn = {9798400706974},
publisher = {Association for Computing Machinery},
address = {New York, NY, USA},
url = {https://doi.org/10.1145/3666025.3699374},
doi = {10.1145/3666025.3699374},
booktitle = {Proceedings of the 22nd ACM Conference on Embedded Networked Sensor Systems},
pages = {784–798},
numpages = {15},
location = {Hangzhou, China},
series = {SenSys '24}
}

@inproceedings{10.1145/3544548.3580801,
author = {Zhang, Ruidong and Li, Ke and Hao, Yihong and Wang, Yufan and Lai, Zhengnan and Guimbreti\`{e}re, Fran\c{c}ois and Zhang, Cheng},
title = {EchoSpeech: Continuous Silent Speech Recognition on Minimally-obtrusive Eyewear Powered by Acoustic Sensing},
year = {2023},
isbn = {9781450394215},
publisher = {Association for Computing Machinery},
address = {New York, NY, USA},
url = {https://doi.org/10.1145/3544548.3580801},
doi = {10.1145/3544548.3580801},
booktitle = {Proceedings of the 2023 CHI Conference on Human Factors in Computing Systems},
articleno = {852},
numpages = {18},
location = {Hamburg, Germany},
series = {CHI '23}
}

@inproceedings{10.1145/3706599.3721185,
author = {Hiraki, Hirotaka and Rekimoto, Jun},
title = {SilentWhisper: inaudible faint whisper speech input for silent speech interaction},
year = {2025},
isbn = {9798400713958},
publisher = {Association for Computing Machinery},
address = {New York, NY, USA},
url = {https://doi.org/10.1145/3706599.3721185},
doi = {10.1145/3706599.3721185},
booktitle = {Proceedings of the Extended Abstracts of the CHI Conference on Human Factors in Computing Systems},
articleno = {746},
numpages = {6},
location = {
},
series = {CHI EA '25}
}

@ARTICLE{Wang2022_TASLP_ACBCFusion,
  author={Wang, Heming and Zhang, Xueliang and Wang, DeLiang},
  journal={IEEE/ACM Transactions on Audio, Speech, and Language Processing}, 
  title={Fusing Bone-Conduction and Air-Conduction Sensors for Complex-Domain Speech Enhancement}, 
  year={2022},
  volume={30},
  number={},
  pages={3134-3143},
  doi={10.1109/TASLP.2022.3209943}}

@article{Wang2022_AppliedAcoustics_MMINet,
  author  = {Mou Wang and Junqi Chen and Xiaolei Zhang and Zhiyong Huang and Susanto Rahardja},
  title   = {Multi-modal speech enhancement with bone-conducted speech in time domain},
  journal = {Applied Acoustics},
  year    = {2022},
  volume  = {200},
  pages   = {109058},
  doi     = {10.1016/j.apacoust.2022.109058}
}

@article{Yu2020_SPL_BoneAir_TimeDomain,
  author  = {Cheng Yu and Kuo-Hsuan Hung and Syu-Siang Wang and Szu-Wei Fu and Yu Tsao and Jeih-Weih Hung},
  title   = {Time-Domain Multi-modal Bone/air Conducted Speech Enhancement},
  journal = {IEEE Signal Processing Letters},
  year    = {2020},
  volume  = {27},
  pages   = {1035--1039},
  doi     = {10.1109/LSP.2020.3000968}
}

@article{Zhou2020_Sensors_BC_RealTime,
  author  = {Yi Zhou and Yufan Chen and Yongbao Ma and Hongqing Liu},
  title   = {A Real-Time Dual-Microphone Speech Enhancement Algorithm Assisted by Bone Conduction Sensor},
  journal = {Sensors},
  year    = {2020},
  volume  = {20},
  number  = {18},
  pages   = {5050},
  doi     = {10.3390/s20185050}
}

@inproceedings{He2023_MobiSys_VibVoice,
  author    = {Lixing He and Haozheng Hou and Shuyao Shi and Xian Shuai and Zhenyu Yan},
  title     = {Towards Bone-Conducted Vibration Speech Enhancement on Head-Mounted Wearables},
  booktitle = {Proceedings of the 21st Annual International Conference on Mobile Systems, Applications, and Services (MobiSys '23)},
  address   = {Helsinki, Finland},
  month     = {June},
  year      = {2023},
  publisher = {ACM},
  doi       = {10.1145/3581791.3596832},
  numpages  = {14}
}

@article{Huang2024_SignalProcessing_OnlineFusion,
  author  = {Boyan Huang and Baiyu Liu and Shuai Zhang and Zhijun Zhang and Tao Zhang and Wenqi Jia and Shiming Zhang and Yifeng Lin and Tetsuya Shimamura},
  title   = {Online bone/air-conducted speech fusion in the presence of strong narrowband noise},
  journal = {Signal Processing},
  year    = {2024},
  volume  = {225},
  pages   = {109615},
  doi     = {10.1016/j.sigpro.2024.109615}
}

@article{Kuang2024_JASA_LightweightFusion,
  author  = {Kelan Kuang and Feiran Yang and Jun Yang},
  title   = {A lightweight speech enhancement network fusing bone- and air-conducted speech},
  journal = {The Journal of the Acoustical Society of America},
  year    = {2024},
  volume  = {156},
  number  = {2},
  pages   = {1355--1366},
  doi     = {10.1121/10.0028339}
}

@standard{ITU-T-P862-2001,
  title        = {Perceptual evaluation of speech quality (PESQ): An objective method for end-to-end speech quality assessment of narrow-band telephone networks and speech codecs},
  organization = {International Telecommunication Union},
  type         = {ITU-T Recommendation},
  number       = {P.862},
  address      = {Geneva, Switzerland},
  year         = {2001}
}

@article{DEMAND,
author={Thiemann, J. and  Ito, N. and  Vincent, E.},
title={{DEMAND}: a collection of multi-channel recordings of acoustic noise in diverse environments},
journal={21st International Congress on Acoustics (ICA 2013)},
url={https://doi.org/10.5281/zenodo.1227121},
year=2013
}

@article{10.1145/3287058,
author = {Maruri, H\'{e}ctor A. Cordourier and Lopez-Meyer, Paulo and Huang, Jonathan and Beltman, Willem Marco and Nachman, Lama and Lu, Hong},
title = {V-Speech: Noise-Robust Speech Capturing Glasses Using Vibration Sensors},
year = {2018},
issue_date = {December 2018},
publisher = {Association for Computing Machinery},
address = {New York, NY, USA},
volume = {2},
number = {4},
url = {https://doi.org/10.1145/3287058},
doi = {10.1145/3287058},
journal = {Proc. ACM Interact. Mob. Wearable Ubiquitous Technol.},
month = dec,
articleno = {180},
numpages = {23}
}

@inproceedings{Rix2001PESQ,
  author    = {Rix, Antony W. and Beerends, John G. and Hollier, Michael P. and Hekstra, Andries P.},
  title     = {Perceptual Evaluation of Speech Quality (PESQ)—A New Method for Speech Quality Assessment of Telephone Networks and Codecs},
  booktitle = {Proceedings of the IEEE International Conference on Acoustics, Speech, and Signal Processing (ICASSP)},
  year      = {2001},
  volume    = {2},
  pages     = {749--752},
  doi       = {10.1109/ICASSP.2001.941023}
}

@article{Taal2011STOI,
  author  = {Taal, Cees H. and Hendriks, Richard C. and Heusdens, Richard and Jensen, Jesper},
  title   = {An Algorithm for Intelligibility Prediction of Time--Frequency Weighted Noisy Speech},
  journal = {IEEE Transactions on Audio, Speech, and Language Processing},
  year    = {2011},
  volume  = {19},
  number  = {7},
  pages   = {2125--2136},
  doi     = {10.1109/TASL.2011.2114881}
}

@inproceedings{Taal2010STOI,
  author    = {Taal, Cees H. and Hendriks, Richard C. and Heusdens, Richard and Jensen, Jesper},
  title     = {A Short-Time Objective Intelligibility Measure for Time-Frequency Weighted Noisy Speech},
  booktitle = {Proc. IEEE Int. Conf. Acoustics, Speech and Signal Processing (ICASSP)},
  year      = {2010},
  pages     = {4214--4217},
  doi       = {10.1109/ICASSP.2010.5495701}
}

@article{moon1990,
author={Moon, J.},
title={The influence of nasal patency on accelerometric transduction of nasal bone vibration},
journal={The Cleft palate journal},
volume=27,
number=3, 
pages={266–274},
doi={10.1597/1545-1569(1990)027<0266:tionpo>2.3.co;2},
year=1990
}

@article{Yiu2012,
author={Yiu, E. M. and  Chen, F. C. and Lo, G. and and Pang, G.},
year=2012,
title={Vibratory and perceptual measurement of resonant voice},
journal={Journal of voice},
volume=26,
number=5,
doi={10.1016/j.jvoice.2012.02.005},
}

@article{kitamura2012,
title={Measurement of vibration velocity pattern of facial surface during phonation using scanning vibrometer},
author={Tatsuya Kitamura},
journal={Acoustical Science and Technology},
volume=33,
number=2,
pages={126-128},
year=2012
}

@misc{freeSTAmerican,
title={ST-AEDS-20180100\_1 Free ST American English Corpus},
howpublished={\url{https://openslr.org/45/}},
author={surfing.ai},
year={2018},
}

@misc{farhadipour2024leveragingselfsupervisedmodelsautomatic,
      title={Leveraging Self-Supervised Models for Automatic Whispered Speech Recognition}, 
      author={Aref Farhadipour and Homa Asadi and Volker Dellwo},
      year={2024},
      eprint={2407.21211},
      archivePrefix={arXiv},
      primaryClass={eess.AS},
      url={https://arxiv.org/abs/2407.21211}, 
}

@inproceedings{10.1145/3544548.3580706,
author = {Rekimoto, Jun},
title = {WESPER: Zero-shot and Realtime Whisper to Normal Voice Conversion for Whisper-based Speech Interactions},
year = {2023},
isbn = {9781450394215},
publisher = {Association for Computing Machinery},
address = {New York, NY, USA},
url = {https://doi.org/10.1145/3544548.3580706},
doi = {10.1145/3544548.3580706},
booktitle = {Proceedings of the 2023 CHI Conference on Human Factors in Computing Systems},
articleno = {700},
numpages = {12},
location = {Hamburg, Germany},
series = {CHI '23}
}

@misc{radford2022robustspeechrecognitionlargescale,
      title={Robust Speech Recognition via Large-Scale Weak Supervision}, 
      author={Alec Radford and Jong Wook Kim and Tao Xu and Greg Brockman and Christine McLeavey and Ilya Sutskever},
      year={2022},
      eprint={2212.04356},
      archivePrefix={arXiv},
      primaryClass={eess.AS},
      url={https://arxiv.org/abs/2212.04356}, 
}

@misc{mushraJS,
title={mushraJS},
author={jfsantos},
url={https://github.com/jfsantos/mushraJS},
year=2019
}

@misc{mushra,
title={BS.1534 : Method for the subjective assessment of intermediate quality level of audio systems},
author={International Telecommunication Union},
url={https://www.itu.int/rec/R-REC-BS.1534/en},
year=2013
}

@misc{V2S200d,
author={Syntiant},
title={{V2S} Voice Vibration Sensor},
howpublished={\url{https://www.syntiant.com/v2s}},
year=2024,
}

@misc{SyntiantMEMS,
author={Syntiant},
title={SiSonic Surface Mount MEMS Microphones},
howpublished={\url{https://www.syntiant.com/mems}},
year=2024,
}

@article{journals/corr/HintonVD15,
  author = {Hinton, Geoffrey E. and Vinyals, Oriol and Dean, Jeffrey},
  ee = {http://arxiv.org/abs/1503.02531},
  journal = {CoRR},
  title = {Distilling the Knowledge in a Neural Network.},
  url = {http://dblp.uni-trier.de/db/journals/corr/corr1503.html#HintonVD15},
  volume = {abs/1503.02531},
  year = 2015,
}

@article{10.1145/3749463,
author = {Wang, Lei and Wang, Xingwei and Zhang, Xi and Ma, Xiaolei and Zhang, Yu and Zhang, Fusang and Gu, Tao and Dai, Haipeng},
title = {AccCall: Enhancing Real-time Phone Call Quality with Smartphone's Built-in Accelerometer},
year = {2025},
issue_date = {September 2025},
publisher = {Association for Computing Machinery},
address = {New York, NY, USA},
volume = {9},
number = {3},
url = {https://doi.org/10.1145/3749463},
doi = {10.1145/3749463},
journal = {Proc. ACM Interact. Mob. Wearable Ubiquitous Technol.},
month = sep,
articleno = {133},
numpages = {33}
}

@inproceedings{10.1145/3613904.3642092,
author = {Wang, Xue and Su, Zixiong and Rekimoto, Jun and Zhang, Yang},
title = {Watch Your Mouth: Silent Speech Recognition with Depth Sensing},
year = {2024},
isbn = {9798400703300},
publisher = {Association for Computing Machinery},
address = {New York, NY, USA},
url = {https://doi.org/10.1145/3613904.3642092},
doi = {10.1145/3613904.3642092},
booktitle = {Proceedings of the 2024 CHI Conference on Human Factors in Computing Systems},
articleno = {323},
numpages = {15},
location = {Honolulu, HI, USA},
series = {CHI '24}
}

@article{decodingSSI2025,
author = {Chowdhury, Adiba Tabassum and  Newaz, Mehrin and Saha, Purnata and AbuHaweeleh, Mohannad Natheef and Mohsen, Sara and Bushnaq, Diala and Chabbouh, Malek and Aljindi, Raghad and Pedersen, Shona and Chowdhury, Muhammad E. H.},
year=2025,
title={Decoding silent speech: a machine learning perspective on data, methods, and frameworks},
journal={Neural Computing and Applications},
pages={6995-7013},
volume=37,
number=10,
url={https://doi.org/10.1007/s00521-024-10456-z},
doi={10.1007/s00521-024-10456-z},
}

@misc{kim2016sequencelevelknowledgedistillation,
      title={Sequence-Level Knowledge Distillation}, 
      author={Yoon Kim and Alexander M. Rush},
      year={2016},
      eprint={1606.07947},
      archivePrefix={arXiv},
      primaryClass={cs.CL},
      url={https://arxiv.org/abs/1606.07947}, 
}

@misc{gandhi2023distilwhisperrobustknowledgedistillation,
      title={Distil-Whisper: Robust Knowledge Distillation via Large-Scale Pseudo Labelling}, 
      author={Sanchit Gandhi and Patrick von Platen and Alexander M. Rush},
      year={2023},
      eprint={2311.00430},
      archivePrefix={arXiv},
      primaryClass={cs.CL},
      url={https://arxiv.org/abs/2311.00430}, 
}

@inproceedings{DCCRNinterspeech20,
author = {Yanxin Hu and Yun Liu and Shubo Lv and Mengtao Xing and Shimin Zhang and Yihui Fu and Jian Wu and Bihong Zhang and Lei Xie},
year=2020,
title={DCCRN: Deep Complex Convolution Recurrent Network for Phase-Aware Speech Enhancement},
booktitle={Proc. Interspeech 2020},
pages={2472-2476},
DOI = {10.21437/Interspeech.2020-2537}
}

@misc{assael2016lipnetendtoendsentencelevellipreading,
      title={LipNet: End-to-End Sentence-level Lipreading}, 
      author={Yannis M. Assael and Brendan Shillingford and Shimon Whiteson and Nando de Freitas},
      year={2016},
      eprint={1611.01599},
      archivePrefix={arXiv},
      primaryClass={cs.LG},
      url={https://arxiv.org/abs/1611.01599}, 
}

@misc{prolific,
author={Prolific inc.},
title={Prolific},
url={https://www.prolific.co},
year=2014,
}

@inproceedings{10.1145/3242587.3242603,
author = {Fukumoto, Masaaki},
title = {SilentVoice: Unnoticeable Voice Input by Ingressive Speech},
year = {2018},
isbn = {9781450359481},
publisher = {Association for Computing Machinery},
address = {New York, NY, USA},
doi = {10.1145/3242587.3242603},
booktitle = {Proceedings of the 31st Annual ACM Symposium on User Interface Software and Technology},
pages = {237–246},
numpages = {10},
location = {Berlin, Germany},
series = {UIST '18}
}

@inproceedings{10.1145/3172944.3172977,
author = {Kapur, Arnav and Kapur, Shreyas and Maes, Pattie},
title = {AlterEgo: A Personalized Wearable Silent Speech Interface},
year = {2018},
isbn = {9781450349451},
publisher = {Association for Computing Machinery},
address = {New York, NY, USA},
doi = {10.1145/3172944.3172977},
booktitle = {23rd International Conference on Intelligent User Interfaces},
pages = {43–53},
numpages = {11},
location = {Tokyo, Japan},
series = {IUI '18}
}

@inproceedings{sottovoce,
author = {Kimura, Naoki and Kono, Michinari and Rekimoto, Jun},
title = {SottoVoce: An Ultrasound Imaging-Based Silent Speech Interaction Using Deep Neural Networks},
year = {2019},
isbn = {9781450359702},
publisher = {Association for Computing Machinery},
address = {New York, NY, USA},
url = {https://doi.org/10.1145/3290605.3300376},
doi = {10.1145/3290605.3300376},
booktitle = {Proceedings of the 2019 CHI Conference on Human Factors in Computing Systems},
pages = {1–11},
numpages = {11},
location = {Glasgow, Scotland Uk},
series = {CHI '19}
}

@inproceedings{10.1145/3544548.3581465,
author = {Su, Zixiong and Fang, Shitao and Rekimoto, Jun},
title = {LipLearner: Customizable Silent Speech Interactions on Mobile Devices},
year = {2023},
isbn = {9781450394215},
publisher = {Association for Computing Machinery},
address = {New York, NY, USA},
url = {https://doi.org/10.1145/3544548.3581465},
doi = {10.1145/3544548.3581465},
booktitle = {Proceedings of the 2023 CHI Conference on Human Factors in Computing Systems},
articleno = {696},
numpages = {21},
location = {Hamburg, Germany},
series = {CHI '23}
}

@ARTICLE{Hsu2021-lz,
  title         = "{HuBERT}: {Self-Supervised} Speech Representation Learning
                   by Masked Prediction of Hidden Units",
  author        = "Hsu, Wei-Ning and Bolte, Benjamin and Tsai, Yao-Hung Hubert
                   and Lakhotia, Kushal and Salakhutdinov, Ruslan and Mohamed,
                   Abdelrahman",
  month         =  jun,
  year          =  2021,
  archivePrefix = "arXiv",
  primaryClass  = "cs.CL",
  eprint        = "2106.07447"
}

\end{document}